\documentclass[final,3p,times]{elsarticle}
\usepackage[table]{xcolor}
\usepackage{bm}
\usepackage{capt-of}

\usepackage{tabularx}
\usepackage{epsfig}
\usepackage{colortbl}
\usepackage{amssymb}
 \usepackage{amsthm}
\usepackage{bm}
\usepackage[colorlinks=true,
            linkcolor=blue,
            citecolor=blue,
            urlcolor=blue]{hyperref}
 \usepackage{lineno}
 \usepackage{amsmath}
 \usepackage{graphicx}
 \usepackage{subfigure}
 \usepackage{color}
\usepackage[shortlabels]{enumitem}
\usepackage{mathrsfs}
\usepackage {float}
\usepackage{caption}
\usepackage[linesnumbered,ruled,vlined]{algorithm2e}
\theoremstyle{plain}

\def\'{~\!'}

\theoremstyle{definition}

\theoremstyle{remark}

\numberwithin{equation}{section}

\allowdisplaybreaks[4]

\journal{.}

\begin{document}

\begin{frontmatter}







\title{Application of Fractional Polynomial Modeling Based on Bayesian Criterion in Cerebrovascular Diseases: A Robust Framework for Occlusion Simulation and Topology Reconstruction}
\tnotetext[foundation]{This work is supported by the National Natural Science Foundation of China (No.12461091), North China University of Technology organized scientific research project (No. 2023YZZKY19, No.2024NCUTYXCX104)}
\author[A]{Yu Zhong*}
\author[A]{Luyao Li }
\cortext[cor1]{Corresponding author: zhongyu@ncut.edu.cn} %
\address[A]{College of Science, North China University of Technology, Beijing 100144, P.R. China}

\begin{abstract}
Stroke is an acute cerebrovascular disease, characterized by high incidence, high mortality, high disability rate and high recurrence rate. The internal carotid artery (ICA), particularly the cervical segment (ICA-C1), is important in cerebrovascular diagnosis and treatment. Its geometry, especially tortuosity, aids disease assessment and surgical planning. Conventional polynomial fitting methods often face order selection, overfitting, and oscillation issues. 

To address these issues, this paper proposes a fractional-order polynomial fitting model based on effective order for morphological analysis of the ICA-C1 segment. A fractional polynomial centerline fitting framework is first constructed to enhance the representation of complex vascular morphology. Then, an automatic order selection mechanism based on the Bayesian Information Criterion (BIC) is introduced to adaptively determine the model order and avoid experience-based manual selection. Furthermore, the concept of “effective order” is proposed, and high-frequency effective orders are identified through statistical analysis of sample data, thereby reducing the candidate order space and significantly improving computational efficiency.

Experimental results on 379 clinical cases show that the proposed method outperforms traditional methods in fitting accuracy, noise robustness, and computational efficiency, achieving low fitting errors and accurately characterizing the complex spatial morphology of the ICA-C1 segment. The model maintains stable and high-quality fitting results under different noise levels (0, 0.01, 0.05 and 0.1), demonstrating strong robustness to data perturbations. Optimization of the solution strategy and order selection mechanism significantly reduced the algorithm's running time from 153.145s to 23.054s while maintaining high fitting stability and accuracy and reducing computational complexity. Furthermore, the proposed model shows promising potential for predicting missing vascular segments in imaging tasks. The prediction results are generally stable, with low errors in most cases, and the normalized mean square error (NMSE) remains below 1.68\% for 90\% of the cases, indicating its potential clinical applicability.

This study provides a stable, efficient, and clinically interpretable modeling method for the precise quantitative analysis of cerebral vascular geometry, providing a new technological and advanced path for intelligent diagnosis and surgical decision-making of cerebrovascular diseases.

\end{abstract}


\end{frontmatter}

\section{Introduction}

\subsection{Research Background }\vskip1mm 

In recent years, with the aging of the population and the change of lifestyle, the incidence rate of cardiovascular and cerebrovascular diseases has continued to rise, which has become one of the main causes of death and disability worldwide \textsuperscript{\cite{feigin2021global, venketasubramanian2017stroke, vos2020global}}. Early risk assessment and precise intervention have become important research priorities in clinical studies. The internal carotid artery (ICA) is an important channel for blood supply to the brain. According to clinical classification standards, ICA is usually divided into multiple anatomical segments, among which segment C1 (ICA–C1) plays an important role in the pathogenesis and surgical pathway of cerebrovascular diseases \textsuperscript{\cite{wolman2022anatomy}}. As an important hub structure connecting the cardiovascular system and cerebral circulation, the morphological changes of the ICA–C1 segment are considered an important anatomical basis for the occurrence of various cardiovascular and cerebrovascular events\textsuperscript{\cite{bridio2021impact}}. The morphological abnormalities of the ICA–C1 segment are closely related to the formation of aneurysms, arterial stenosis, ischemic stroke, and complications of interventional surgery\textsuperscript{\cite{ulus2022anatomical}}.

The blood vessels in the ICA–C1 segment can be categorized according to their tortuosity patterns into four common types: straight, tortuous, coiled, and kinked. The straight type is defined when the angle between the centerlines of the common carotid artery and the internal carotid artery is less than 15°. The tortuous type exhibits a smooth bending morphology resembling “C”, “U”, or “S” shapes. The coiled type forms one or more helical loops along its longitudinal axis. The kinked type features sharp angular bends, often forming a “V”-shaped kink\textsuperscript{\cite{koge2022internal}}. Representative examples are illustrated in Figure 1.

\begin{figure}[H]  
  \centering
  \includegraphics[width=0.9\textwidth]{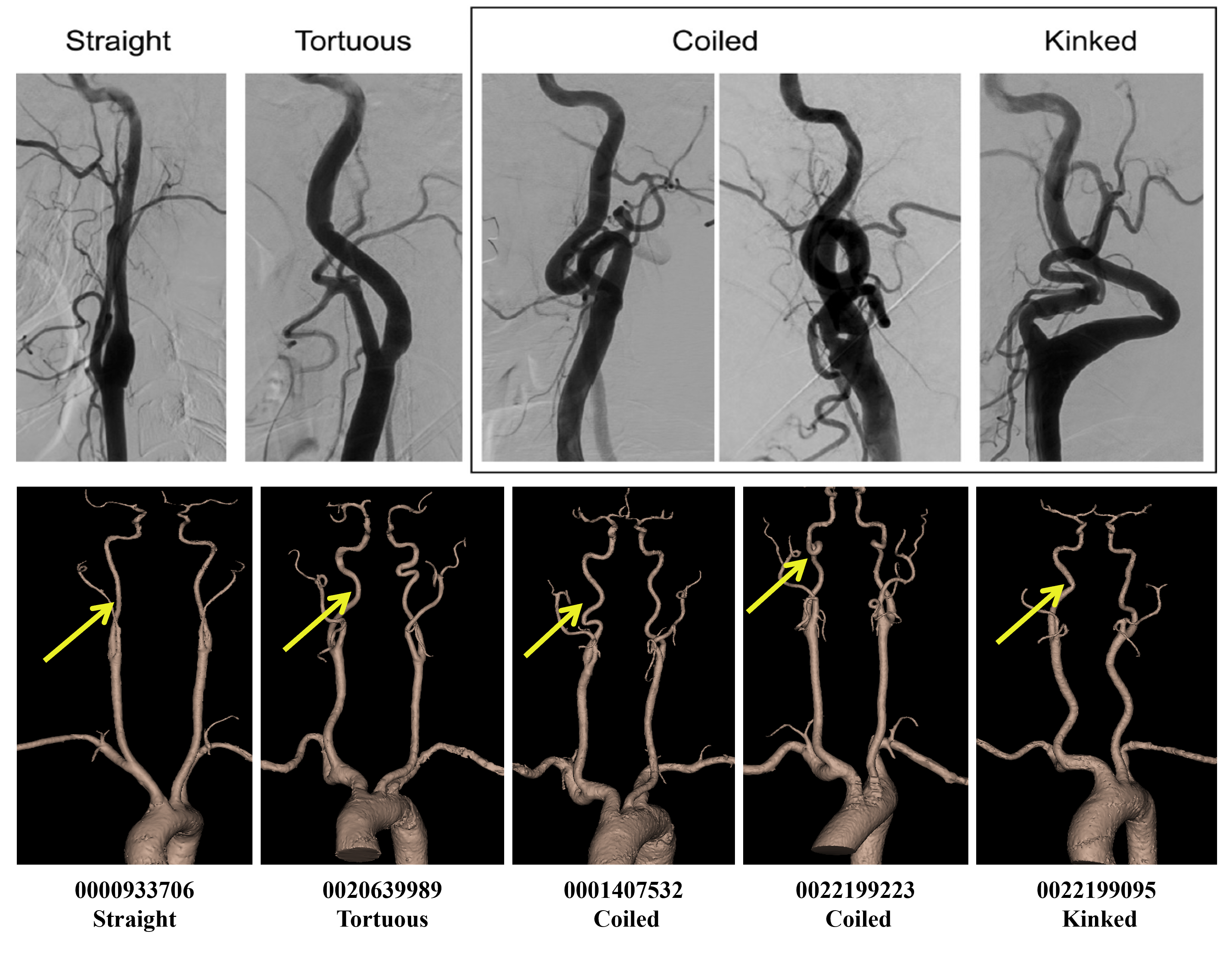} 
  \caption{Different types of digital subtraction angiography images of ICA-C1 segment (in literature \cite{koge2022internal}) and three-dimensional vascular schematic diagram of ICA-C1 segment reconstructed using Mimics}
  \label{fig:example}
\end{figure}

In recent years, vascular morphological indicators, especially vascular tortuosity, have gradually become important parameters for quantitative analysis of cerebrovascular diseases\textsuperscript{\cite{Deshpande2022Novel,Sun2022AgeRelated}}. By accurately modeling the centerline of blood vessels, curvature, torsion, and related index can be further calculated to quantitatively describe the spatial morphological changes of blood vessels\textsuperscript{\cite{Piccinelli2009A}}. These geometric parameters not only contribute to disease classification and risk assessment, but also have important application value in preoperative path planning and catheter navigation\textsuperscript{\cite{Decroocq2022Modeling}}.

Traditional vascular morphology analysis methods face two main problems. Firstly, although polynomial or B-spline fitting and other analytical methods have high accuracy, they are sensitive to high-order derivatives and easily affected by fitting errors\textsuperscript{\cite{Decroocq2022Modeling}}. The second reason is that the calculation method based on discrete points has low robustness and is easily affected by noise interference, resulting in unstable estimates of curvature and torsion\textsuperscript{\cite{Tyrrell2007Robust}}. Therefore, how to balance expression ability, computational stability, and efficiency in the centerline modeling method has become an urgent problem to be solved.

\subsection{Research status}\vskip1mm
At present, there are mainly two types of methods for calculating the curvature and torsion of blood vessel centerlines. One type is based on curve fitting, which mainly fits an approximate smooth curve at the target point and its neighboring points, and then uses the differential geometry formula of the analytical curve to calculate the curvature and torsion. Typical methods include polynomial fitting\textsuperscript{\cite{ostertagova2012modelling}} and B-spline fitting\textsuperscript{\cite{schoenberg1946contributions}}. Its advantage is that it can obtain analytical function expressions and achieve high computational accuracy by analytically differentiating approximate curves. In vascular modeling and morphological analysis, there have been numerous application cases. For example, Kobayashi et al. \textsuperscript{\cite{Kobayashi2020A}}used five penalty spline fitting to fit the centerline of the blood vessel, obtained high-precision curvature and torsion through analytical differentiation, and achieved three-dimensional morphological quantification of the abdominal aortic aneurysm stent and internal carotid artery over time. Michael J. Johnson et al. \textsuperscript{\cite{johnson2007robust}}proposed a three-dimensional vascular morphology quantification method based on minimum curvature. This method obtains the tortuosity index of blood vessels by fitting the centerline of the vessel with polynomial spline and calculating the three-dimensional curvature. Chakravarty A. et al. \textsuperscript{\cite{chakravarty2013novel}}proposed a method based on quadratic polynomial decomposition for quantitatively evaluating the tortuosity of retinal blood vessels. This method decomposes the vascular curve into multiple quadratic polynomial segments, calculates the curvature changes of each segment, and obtains the tortuosity index of the blood vessels. Yo Ping H. et al. \textsuperscript{\cite{huang2024computer}}used polynomial curve fitting to enhance the ability to measure the degree of retinal vascular tortuosity and achieved good results. Polynomial fitting can often provide analytical function expressions, and the differentiation of analytical functions can provide higher accuracy, but it is often highly sensitive to fitting errors due to its dependence on high-order derivative calculations. 
 
Another type is based on discrete methods, which do not require fitting analytical curves, but directly use the geometric relationships of discrete points to estimate curvature and torsion. Typical methods include projection method\textsuperscript{\cite{mardia1999estimation}}, unilateral difference method\textsuperscript{\cite{blankenburg2016parameter}}, and discrete geometry method\textsuperscript{\cite{an2011geometric}}. When evaluating the tortuosity of the aorta, Zhang et al. \textsuperscript{\cite{zhang2021application}}applied the micro center difference algorithm based on the 3D centerline to obtain the curvature and torsion, thereby obtaining the tortuosity characteristics of the blood vessel. Alexander B. B. et al. \textsuperscript{\cite{brummer2020improving}}applied the fifth order central difference approximation to calculate the curvature torsion of the vessel centerline, and derived the vessel morphology based on the Frenet-Serret equation with high accuracy. An et al. \textsuperscript{\cite{an2020geometric}}proposed a curvature and torsion estimation method based on Gaussian weighted discrete derivatives, which directly estimates tangential, normal, curvature, and torsion from one-dimensional point clouds, and improves robustness through noise adaptive neighborhood control. However, the above methods often use discrete point approximation to calculate curvature torsion, which increases the response of the calculation method to noise and may result in significant errors.

To solve the above problems, fractional order modeling theory\textsuperscript{\cite{Royston1994}} provides a flexible new approach, which allowing polynomial orders to be non-integer and more continuous, achieving fine adjustment between integer orders, enhancing adaptability to complex vascular morphology and alleviating high-order oscillations. Based on this, fractional polynomial centerline modeling combined with data-driven order optimization mechanism has become a new direction for vascular morphology research in the ICA–C1 segment.

\subsection{Outline}\vskip1mm 
\begin{figure}[H]  
  \centering
  \includegraphics[width=1\textwidth]{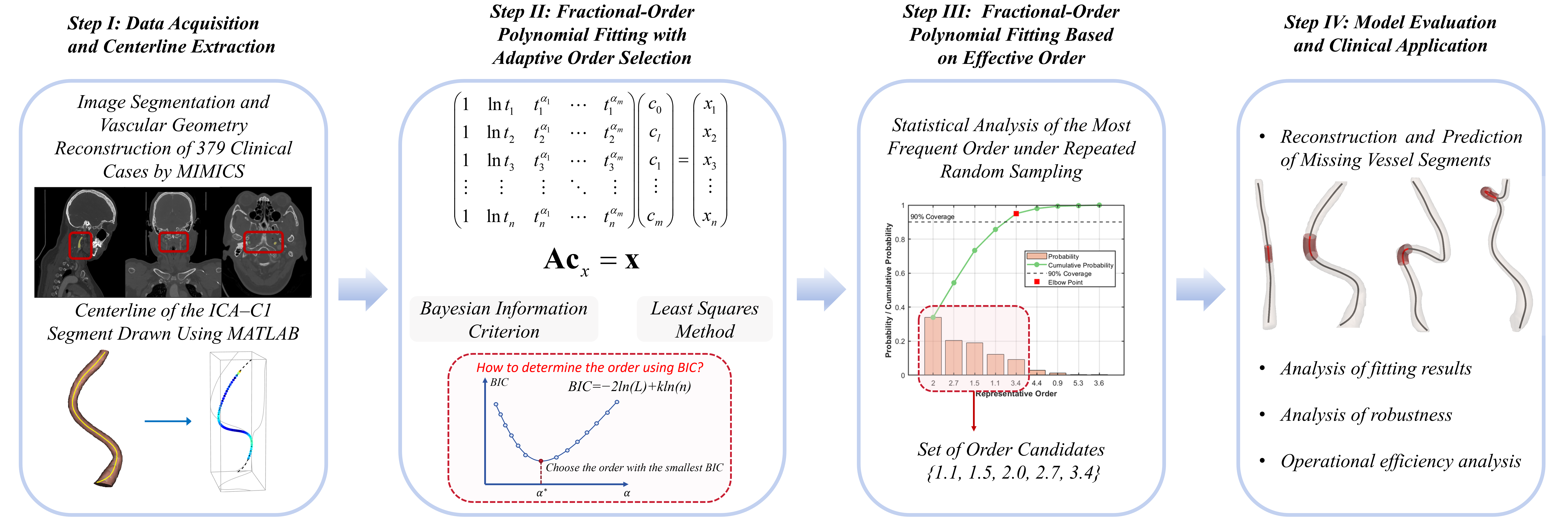} 
  \caption{Workflow of fractional-order polynomial modeling, BIC-based order selection, and accelerated strategy for ICA-C1 centerline reconstruction and validation}
  \label{fig:example}
\end{figure}
This study focuses on the geometric modeling of the centerline of the ICA–C1 segment.A novel fractional polynomial fitting method based on effective order is proposed and validated using large-scale clinical data. The remainder of this paper is organized as follows.

In Section 2, the research methodology is presented. The data sources and the methods for image-based 3D geometric reconstruction and vascular centerline extraction are first introduced. Based on this, a fractional-order polynomial centerline fitting model is constructed to better represent complex vascular morphology. An automatic order selection mechanism based on the Bayesian Information Criterion (BIC)\textsuperscript{\cite{burnham2004multimodel}} is then adopted to adaptively determine the model order and avoid uncertainty caused by manual settings. Furthermore, the concept of “effective order” is proposed, and high-frequency effective orders are identified through statistical analysis of 100 samples to reduce the algorithm search space, with the improved algorithm workflow and implementation details also provided.

In Section 3, the experimental results and analysis are presented. Based on 379 clinical cases, the improved algorithm was evaluated in terms of fitting accuracy, computational efficiency, and noise robustness, with geometric indicators used to analyze the tortuosity characteristics of the ICA–C1 segment.The results demonstrate high fitting accuracy, strong robustness under noise levels of 0-0.1, and a significant reduction in running time from 153.145s to 23.054s. In predicting missing vascular segments, 90\% of cases achieved an NMSE below 1.68\%, indicating promising clinical application value.

In Section 4, the research findings are discussed. This section analyzes the advantages and limitations of the proposed method, and looks forward to future improvement directions and potential clinical applications. The entire work and main research conclusions are summarized.

\section{\bf Methodology}

\subsection{Data source and preprocessing}\vskip1mm 

The data for this study was collected from Beijing Chaoyang Hospital affiliated to Capital Medical University from March to November 2024. The clinical data was stored anonymously, and all patients were informed and agreed to participate in the aforementioned study. The application of the study's data has been reviewed by the ethics committee and a confidentiality agreement has been signed. The CTA scan images of all participants were obtained in DICOM format and processed using the medical professional software Mimics to reconstruct the three-dimensional geometric model of the ICA-C1 segment. Subsequently, the reconstructed vessels were smoothed using the Laplacian smoothing method to reduce noise caused by image data discontinuity or uncertainty in lumen boundary identification. Finally, the centerline was discretized into a series of discrete points, with a number of 45-120, to characterize the geometric features of the ICA-C1 segment vessel centerline for subsequent vascular tortuosity calculation.

\subsection{Fractional polynomial centerline fitting}

\subsubsection{Spatial curve and fractional polynomial}\vskip1mm 
\paragraph{(1) Three-dimensional curvature and torsion}\mbox{}

Blood vessels exhibit complex meandering and tortuous structures in three-dimensional space. Precise quantification of their morphology serves as the foundation for understanding hemodynamics, assessing pathological conditions, and planning surgeries\textsuperscript{\cite{Bullitt2003Measuring, zhang2021application}}. In the research process, the three-dimensional morphology of blood vessels can be abstracted as a spatial curve formed by its centerline. Curvature and torsion provide the basis for model research.

In differential geometry, a continuously differentiable curve \(\mathbf{r}(s)\) in three-dimensional space is usually parameterized by arc length \(s\). Its local shape can be characterized by two geometric invariants, which uniquely describe its local geometry, i.e., curvature \(\kappa\) and torsion \(\tau\).

The curvature \(\kappa(s)\) of a curve at a certain point quantifies the degree of its deviation from a straight line, that is, the degree of bending of the curve. Its mathematical definition is the magnitude of the rate of change of the unit tangent vector \(\mathbf{T}(s)\) of the curve with respect to the change in arc length.
$
\kappa(s) = \left\| \frac{d\mathbf{T}(s)}{ds} \right\| = \| \mathbf{T}'(s) \| \geq 0.
$
Among them, \(\mathbf{T}(s) = \mathbf{r}'(s)\). Zero curvature indicates that the vicinity of the point is a straight line segment, meaning it does not exhibit any bending characteristics. The torsion \(\tau(s)\) of a curve at a certain point quantifies the degree of deviation from a planar curve, that is, the degree of twisting of the curve. It describes the rate of change in the direction of the local osculating plane of the curve. It is defined as
$
\tau(s) = -\mathbf{N}(s) \cdot \frac{d\mathbf{B}(s)}{ds} = -\mathbf{N}(s) \cdot. \mathbf{B}'(s)
$
where \(\mathbf{N}(s)\) is the unit principal normal vector (\(\mathbf{N} = \mathbf{T}' / \|\mathbf{T}'\|\)),  \(\mathbf{B}(s)\) is the unit binormal vector (\(\mathbf{B} = \mathbf{T} \times \mathbf{N}\)), and these three vectors constitute the Frenet frame \(\{\mathbf{T}, \mathbf{N}, \mathbf{B}\}\) describing the local direction of the curve. The torsion can be positive or negative, and its sign indicates the direction of twist. Zero torsion means that the curve near that point lies in a plane.

In practical applications, the centerline of blood vessels is usually given in the form of discrete points, and curvature and torsion are generally calculated based on parameterized spatial curves. If the centerline is represented as a three-dimensional parameter curve \(\mathbf{r}(t)\) (\(t\) is any parameter), then \(\kappa\) and \(\tau\) can be given by the following equations.

\begin{equation}
\left\{
\begin{aligned}
\kappa(t) &= \left|\frac{d\boldsymbol{T}(s)}{ds}\right|
= \frac{ \left\| \bm{r}'(t) \times \bm{r}''(t) \right\| }
{ \left\| \bm{r}'(t) \right\|^3 }, \\[6pt]
\tau(t) &= -\frac{d\mathbf{B}(s)}{ds} \cdot \mathbf{N}(s)
= \frac{ \left( \bm{r}'(t) \times \bm{r}''(t) \right) \cdot \bm{r}'''(t) }
{ \left\| \bm{r}'(t) \times \bm{r}''(t) \right\|^2 } .
\end{aligned}
\right.
\end{equation}

According to the fundamental theorems of differential geometry, a spatial curve is uniquely determined by its curvature function \(\kappa(t)\) and torsion function \(\tau(t)\), under the meaning of rigid body motion (translation and rotation). Therefore, the three-dimensional shape of the vascular centerline can be fully described by these two indicators\textsuperscript{\cite{Shabana2022Curvature, Ross2019The}}.

Curvature and torsion, as core differential geometric invariants of three-dimensional space curves, provide a rigorous, universal, and comprehensive mathematical language for describing the morphology of vascular centerlines, transforming complex visual forms into computable and analyzable functions.

\paragraph{(2) Theory of fractional polynomial model}\mbox{}

There have been long-standing issues in the field of vascular centerline fitting research, such as the singular method for describing vascular tortuosity and the reliance on subjective judgment for vascular classification\textsuperscript{\cite{Kobayashi2020A}}. To address the limitations of describing vascular tortuosity, this paper introduces the concept of fractional-order polynomials and optimizes them based on this. Fractional-order polynomials were first proposed by Royston and Altman et al. \textsuperscript{\cite{Royston1994}}, with the core idea of extending the power of integer-order polynomials to the fractional domain.

Fractional polynomials have significant advantages in the field of high-precision regression modeling. On the one hand, they can represent more complex local geometric features with lower orders, reducing computational load while minimizing the amplification effect of higher orders on noise signals\textsuperscript{\cite{Sauerbrei2006Multivariable, Regier2015Smoothing}}. On the other hand, they inherit the accuracy advantages of analytical fitting methods, enabling stable curvature and torsion calculations in analytical forms, which has obvious advantages in vascular morphology applications. Generally speaking, the \(m\)-order integer polynomial regression model is

\begin{equation}
    y=\beta_{0}+\beta_{1} x+\beta_{2} x^{2}+\cdots+\beta_{m} x^{m}=\beta_{0}+\sum_{j=1}^{m} \beta_{j} x^{j} \quad \forall j=1,2, \cdots, m
\end{equation}

If the range of power exponents is extended from positive integers to the fractional domain, fractional-order polynomials can be defined. Correspondingly, according to Royston's FP model, the fractional-order polynomial regression model can be expressed as

\begin{equation}
   FP(x)=\beta_{0}+\sum_{j=1}^{m} \beta_{j} x^{\left(\alpha_{j}\right)} \quad \forall j=1,2, \cdots, m
\end{equation}

where
\[
x^{(\alpha_j)} =
\begin{cases}
x^{\alpha_j}, & \alpha_j \neq 0, \\
\ln x, & \alpha_j = 0
\end{cases}
\]

Here, $\alpha_1 < \alpha_2 < \dots < \alpha_m$, each $\alpha_j$ represents a fractional order. If  $\alpha_1 \leq \alpha_2 \leq \dots \leq \alpha_m$ and repeated powers occur, the basis functions are modified by introducing a $\ln x$, forming Royston's FP2 model. In this study, fractional polynomial is mainly used for geometric fitting and modeling of the ICA-C1 segment. Given the relatively straight and limited complexity of this vascular segment, to simplify the calculation, this paper adopts Royston's FP1 model as a reference and employs a fixed step size incremental power selection strategy to construct a power set, i.e., $\alpha_1 < \alpha_2 < \dots < \alpha_m$, as the basis for the fitting model in this study.

In existing research, the fractional polynomial fitting model relies on the manually predefined highest order number $\alpha_{\max}$. However, the value of $\alpha_{\max}$ simultaneously affects both fitting accuracy and computational complexity. When the order is low, the model's expressive power is insufficient, leading to underfitting. When the order is too high, it introduces more parameters to be estimated, resulting in increased computational overhead. To achieve automatic order selection and balance model accuracy and computational efficiency, this paper introduces the Akaike Information Criterion (AIC) and Bayesian Information Criterion (BIC) to evaluate candidate $\alpha_{\max}$ values, and selects the order corresponding to the optimal information criterion as the final model order.

\subsubsection{Determination of fractional polynomial coefficients }\vskip1mm 
The 3D coordinate data of the sample is derived from discrete centerline points $\{(x_i,y_i,z_i)\}_{i=1}^{N}$. Based on this, cumulative chord length parameterization is performed and fractional polynomial fitting is implemented.

First, calculate the chord length increment between adjacent points as $ds_i=\sqrt{(x_{i+1}-x_i)^2+(y_{i+1}-y_i)^2+(z_{i+1}-z_i)^2},\\ \quad i=1,\dots,N-1$, to obtain the cumulative chord length parameter $s_1=0,\quad s_i=\sum_{j=1}^{i-1}ds_j,\quad i=2,\dots,N$. To eliminate the influence of differences in the length of centerlines among different samples, the chord length should be normalized into a dimensionless parameter

\begin{equation}
t_i=\frac{s_i-\min(\mathbf{s})}{\max(\mathbf{s})-\min(\mathbf{s})}+0.01,
\end{equation}

It is worth noting that to avoid the numerical singularity issue of $\ln 0 $ at $ t = 0$, a small value $\epsilon$ should be introduced to shift the entire graph to the interval $[\epsilon, 1 + \epsilon]$, namely [0.01, 1.01]. This operation does not affect the geometric characteristics of the curve. Based on this, we can construct a fractional-order polynomial fitting model that includes constant terms, fractional power terms, and logarithmic terms.

For a given maximum fractional order $\alpha_{\max}$, the fractional order power parameter $\alpha$ takes values of $0.1, 0.2, \dots, \alpha_{m}$ with a step size of 0.1, satisfying the requirement that the fractional polynomial orders $\alpha_1 < \alpha_2 < \dots < \alpha_m$. For each sample, to avoid manually specifying $\alpha_{\max}$, this paper sets the candidate set of highest orders as $\alpha_{\max} \in \{ 0.1, 0.2, \dots, 0.1(N-2) \}$, where N is the number of discrete points on the centerline, with a step size of 0.1. For each candidate $\alpha_{\max}$, let $\alpha_{m} = \alpha_{\max}$, construct a matrix A consisting of constant terms, fractional power terms of $t$, and $\ln t$ terms, and fit the three-dimensional coordinate components separately.

\begin{equation}
\mathbf{A}=\left[ \mathbf{1},\ t^{0.1},\ t^{0.2},\ \dots,\ t^{\alpha_{\max}},\ \ln(t)\right]\in\mathbb{R}^{N\times p},
\end{equation}

Among them, the number of terms is given by $p=10|\alpha_{\max}|+2$ (including the constant term and the $\ln(t)$ term), with $p \le N$, and $\mathbf{1}$ is an all-1 vector.

Under the conditions that $p \le N$, $t_i$ are distinct, and the matrix of basis functions has full column rank, $A^\mathrm{T}$ is invertible, and the closed-form solution of the least squares method can be given by the normal equation. Therefore, coefficient estimates can be obtained from the least squares method\textsuperscript{\cite{Cetisli2011}}.

\begin{equation}
\left\{
\begin{aligned}
\hat{\mathbf{c}}_x&=\arg\min_{\mathbf{c}}\|\mathbf{x}-\mathbf{A}\mathbf{c}\|^2
=(\mathbf{A}^\mathrm{T}\mathbf{A})^{-1}\mathbf{A}^\mathrm{T}\mathbf{x},\\
\hat{\mathbf{c}}_y&=\arg\min_{\mathbf{c}}\|\mathbf{y}-\mathbf{A}\mathbf{c}\|^2
=(\mathbf{A}^\mathrm{T}\mathbf{A})^{-1}\mathbf{A}^\mathrm{T}\mathbf{y},\\
\hat{\mathbf{c}}_z&=\arg\min_{\mathbf{c}}\|\mathbf{z}-\mathbf{A}\mathbf{c}\|^2
=(\mathbf{A}^\mathrm{T}\mathbf{A})^{-1}\mathbf{A}^\mathrm{T}\mathbf{z}.
\end{aligned}
\right.
\end{equation}

Thus, a fractional-order fitting curve is obtained
\begin{equation}
\left\{
\begin{aligned}
\hat{x}_i &= \hat{c}_{x,0}+\sum_{j=1}^{m}\hat{c}_{x,j}\,t_i^{\alpha_j}+\hat{c}_{x,\ln}\ln(t_i),\\
\hat{y}_i &= \hat{c}_{y,0}+\sum_{j=1}^{m}\hat{c}_{y,j}\,t_i^{\alpha_j}+\hat{c}_{y,\ln}\ln(t_i),\\
\hat{z}_i &= \hat{c}_{z,0}+\sum_{j=1}^{m}\hat{c}_{z,j}\,t_i^{\alpha_j}+\hat{c}_{z,\ln}\ln(t_i),
\end{aligned}
\right.
\quad i=1,\ldots,N.
\end{equation}

That is,

\begin{equation}
\hat{\mathbf{x}}_{\mathrm{frac}}=\mathbf{A}\hat{\mathbf{c}}_x,\quad
\hat{\mathbf{y}}_{\mathrm{frac}}=\mathbf{A}\hat{\mathbf{c}}_y,\quad
\hat{\mathbf{z}}_{\mathrm{frac}}=\mathbf{A}\hat{\mathbf{c}}_z.
\end{equation}

\subsubsection{Adaptive selection of fractional order based on BIC }

From Section 2.2.2, it can be concluded that for each candidate maximum order number $\alpha_{\max}$, a corresponding set of fitting results can be obtained. To quantitatively evaluate the fitting performance of different model orders, several statistical indicators are introduced in this study.

First, the Mean Squared Error (MSE)\textsuperscript{\cite{chicco2021coefficient}} is adopted as the primary metric for measuring fitting accuracy.
$
MSE=\frac{1}{N}\sum_{i=1}^{N}(y_i-\hat{y}_i)^2
$
where $y_i$ denotes the observed value, $\hat{y}_i$ represents the predicted value, and $N$ is the total number of samples. The MSE reflects the average squared deviation between the predicted results and the actual observations. Since the squared operation amplifies larger errors, MSE is particularly sensitive to significant deviations. A smaller MSE value indicates that the predicted curve is closer to the observed data, implying better fitting performance.

However, when datasets have different magnitudes or scales, direct comparison using MSE alone may become unreliable. Therefore, the Normalized Mean Squared Error (NMSE) is further introduced as an auxiliary evaluation index to reduce the influence of scale variations among different samples.
$
\mathrm{NMSE}
=
\frac{
\sum_{i=1}^{N}(y_i-\hat{y}_i)^2
}{
\sum_{i=1}^{N}(y_i-\bar{y})^2
}
$
where $\bar{y}$ is the mean of the observed values. NMSE normalizes the residual error by the variance of the observed data, enabling a more consistent comparison across different datasets or operating conditions. A smaller NMSE value indicates higher prediction accuracy and stronger fitting capability.\textsuperscript{\cite{Li2023Ship}}

Although increasing the model order generally improves fitting accuracy, excessive model complexity may lead to overfitting, thereby reducing the model’s generalization ability. To achieve a balance between fitting accuracy and model complexity, the Akaike Information Criterion (AIC) and Bayesian Information Criterion (BIC)\textsuperscript{\cite{burnham2004multimodel}} are commonly employed for model selection.

The AIC is defined as
$
\text{AIC} = -2\ln(L) + 2k=N \ln\left(\frac{RSS}{N}\right) + 2k + C,
$
where $L$ denotes the likelihood function, $k$ is the number of model parameters, $RSS$ is the residual sum of squares, $N$ represents the sample size, and $C$ is a constant independent of the model structure. The first term evaluates the goodness of fit, while the second term penalizes model complexity. Therefore, AIC attempts to identify a model that achieves good fitting performance with a relatively small number of parameters. Lower AIC values indicate a better trade-off between accuracy and complexity.

Similarly, BIC is defined as
$
\text{BIC} = -2\ln(L) + k \ln(N)=N \ln\left(\frac{RSS}{N}\right) + k \ln(N) + C.
$
Compared with AIC, the penalty term in BIC increases with the sample size $N$, meaning that BIC imposes a stronger penalty on overly complex models. Consequently, BIC generally favors simpler models and is often more effective in avoiding overfitting when the sample size is large. Lower BIC values correspond to models with better overall performance considering both fitting quality and structural simplicity.

In this study, BIC is employed as the primary criterion for determining the optimal model order. Different candidate values of $\alpha_{\max}$ are tested sequentially, and the corresponding BIC values are calculated for each fitted model. The model associated with the minimum BIC value is selected as the optimal order, since it provides the most appropriate balance between fitting accuracy and model complexity. This strategy ensures that the selected model not only captures the essential characteristics of the data but also maintains good computational efficiency.

\renewcommand{\arraystretch}{1.5}
\setlength{\arrayrulewidth}{1pt}
\begin{center}
\captionof{table}{Comparison of AIC/BIC order selection and fit}
\label{tab:AIC BIC}

\begin{tabular}{|c|c|c|c|c|c|}
\hline
\rowcolor{gray!40}
\textbf{ID} & \textbf{Order(AIC)} & \textbf{MSE(AIC)} & \textbf{Order(BIC)} & \textbf{MSE (BIC)} & \textbf{MSE difference} \\
\hline
0022293785\_2 (Straight) & 5.6 & 0.0011 & 2.1 & 0.0038 & 0.0027 \\
\hline
0022255480\_1 (Tortuous) & 3.9 & 0.0033 & 2 & 0.0093 & 0.0060 \\
\hline
0000090587\_2 (Kinked) & 3.4 & 0.0617 & 1.5 & 0.1585 & 0.0968 \\
\hline
0022199223\_1 (Coiled) & 6.5 & 0.0050 & 2.7 & 0.0206 & 0.0156 \\
\hline
\end{tabular}
\end{center}

The reason this paper chooses BIC as the primary model selection criterion is that in the CTA images of this study, generally 45 to 120 vessel centerline points can be extracted. When $n=45$, the BIC penalty term is $k \cdot \ln 45 > 3k$, while the AIC penalty term is $2k$. Thus, under the same sample size, BIC imposes a stricter constraint on model complexity. As shown in Table 1, in the fitting of different types of vessels, the models selected by AIC and BIC are similar in fitting accuracy, but AIC tends to favor selecting higher orders, leading to increased model complexity and computational overhead. BIC, on the other hand, can effectively control the model order while ensuring fitting accuracy, simplifying the structure and reducing the computational burden, which is more in line with the efficiency requirements of clinical data processing. Even for the two most tortuous cases, which also have the largest MSE differences, the discrepancy between the two remains very small (Figure 3). Furthermore, the ICA-C1 segment, as a thick segment of the internal carotid artery, has an average curvature of approximately 0.06 and an average radius of 4.5 to 5.7 mm \textsuperscript{\cite{Baz2021}}. Its morphological characteristics determine that the model allows for a certain degree of fitting error under the premise of accurate overall trend. Therefore, excessively increasing the order does not significantly improve the analysis results but may instead weaken research efficiency due to the increased computational workload.

\begin{figure}[H]  
  \centering
  \includegraphics[width=1\textwidth]{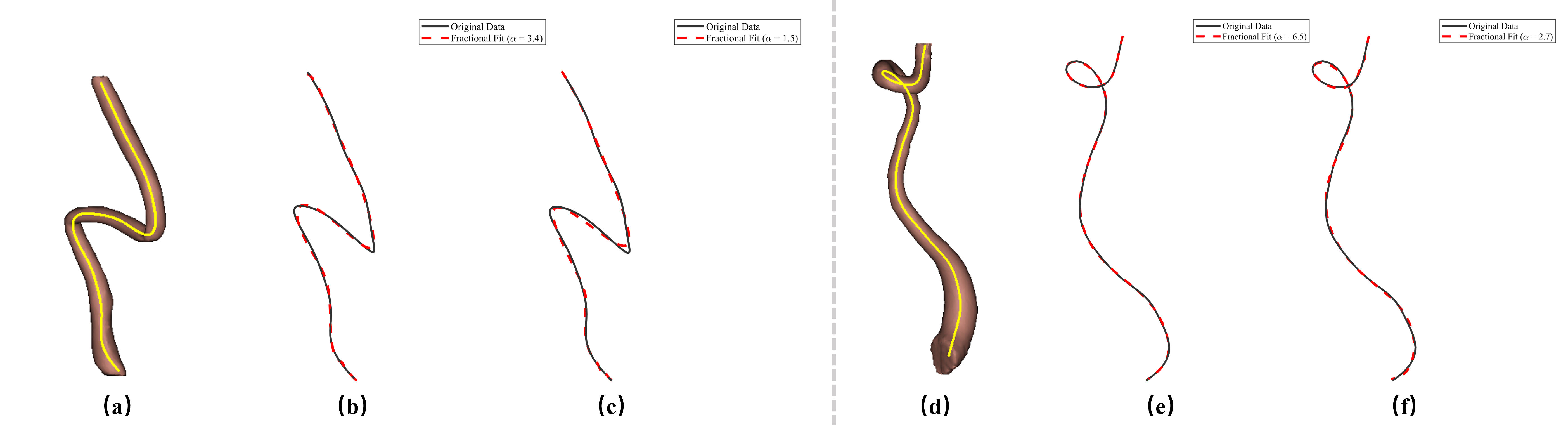} 
  \caption{Comparison of fitting results for the case with the largest MSE difference (0000090587\_2 and 0022199223\_1):
(a)and(d) 3D reconstruction of the ICA–C1 segment;
(b)and(e) Fitting plot based on AIC-selected order;
(c)and(f) Fitting plot based on BIC-selected order.}
  \label{fig:kinked coiled}
\end{figure}

In summary, this paper ultimately selects the order $\alpha_{m}$ corresponding to the minimum BIC as the optimal highest order of the fractional polynomial, ensuring a balance between model accuracy and complexity. At the same time, AIC is employed as an auxiliary reference to verify and enhance the robustness of the selection, providing additional confidence in the chosen model order.

\subsection{Effective order}\vskip1mm 
\subsubsection{Definition and extraction of effective order}
To reduce the computational cost of searching for optimal orders in fractional polynomial fitting, three independent rounds of random sampling were conducted, each selecting 100 samples from the full dataset. For each sample, the optimal maximum order $\alpha_{\max}^{*}$ was automatically selected using the BIC criterion. The frequency of each selected order across the samples is shown in Figure 4(a). 

\begin{figure}[H] 
\centering 
\includegraphics[width=1\textwidth]{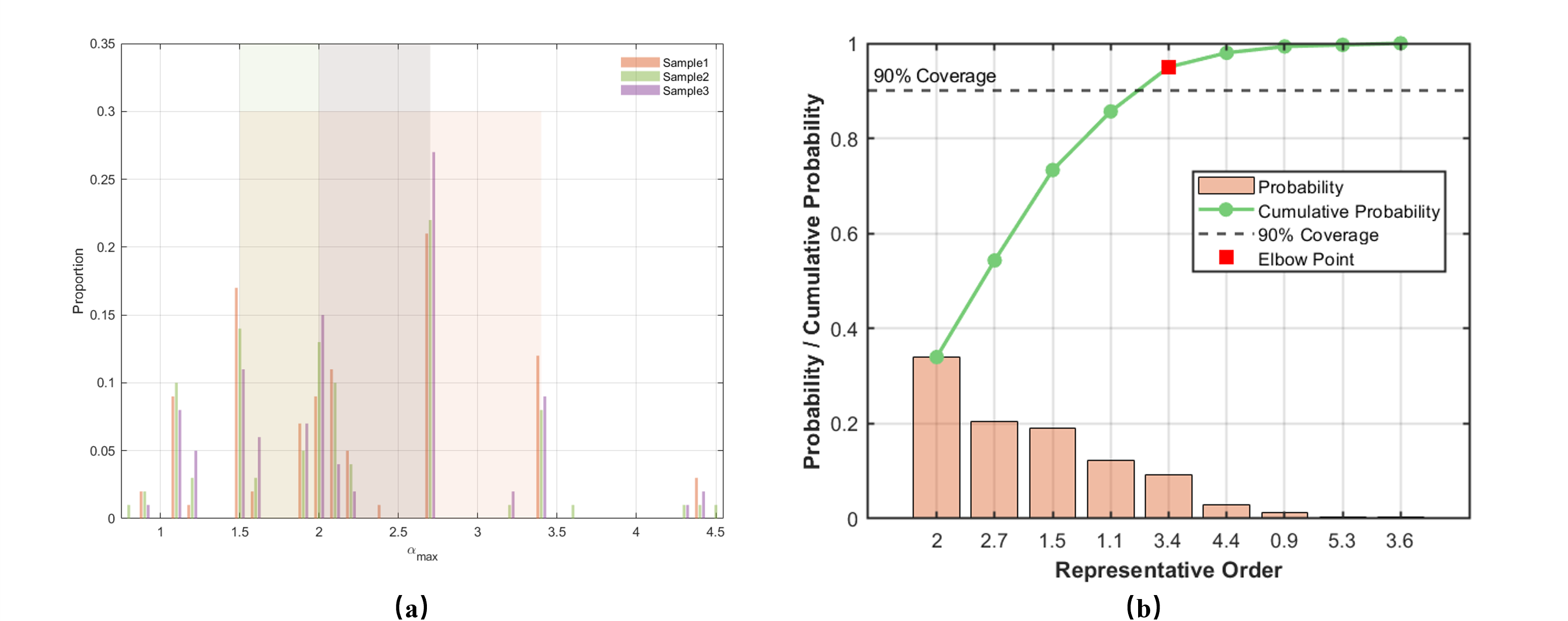} 
\caption{Order distributions and order selection: (a) probability distribution of sample order; (b) Order selection and cumulative probability.} 
\label{fig:example} 
\end{figure}

Certain orders consistently appeared with high frequency across the samples, indicating that these orders reliably capture the typical complexity required to fit the data. Orders and their frequencies form the dataset
$
X = \{(x_i, f_i)\}_{i=1}^{N},
$
where $x_i$ denotes the order and $f_i$ its occurrence frequency. Since some orders are numerically close (e.g., 1.9, 2.0, 2.1, with a step size of 0.1), treating them as distinct would introduce redundancy. Therefore, numerically adjacent orders are merged to extract representative orders summarizing the main trends.

K-means clustering was applied to group similar orders, minimizing intra-cluster distances without manually defining intervals. The optimal number of clusters was determined by maximizing the Silhouette coefficient, balancing intra-cluster compactness and inter-cluster separability:
$
S(k) = \frac{1}{N} \sum_{i=1}^{N} \frac{b_i - a_i}{\max(a_i, b_i)},
$
where $a_i$ is the average intra-cluster distance and $b_i$ the distance to the nearest neighboring cluster.

The representative order of each cluster was taken as the mode:
$
c_j = \operatorname{mode}(C_j),
$
and the cluster weight, quantifying the contribution of each order, was defined as
$
w_j = \sum_{x_i \in C_j} f_i.
$

Figure 4(b) shows the probability distribution and cumulative probability of the representative orders. The final set of effective orders selected for subsequent fractional polynomial modeling is $\{1.1, 1.5, 2.0, 2.7, 3.4\}$. These orders are identified from the region preceding the elbow point, where the cumulative probability curve exhibits a sharp increase before gradually plateauing near 90\%. This behavior indicates that most of the informative contribution is concentrated within a limited number of candidate orders, while the remaining higher-order terms contribute only marginal gains.

By selecting $\{1.1, 1.5, 2.0, 2.7, 3.4\}$, the proposed method effectively preserves the most representative structural characteristics of the underlying distribution while avoiding unnecessary model complexity. In particular, these selected fractional orders provide a balanced coverage of both low- and mid-range nonlinear effects, enabling the model to capture subtle variations in the data without overfitting. Consequently, this reduced and well-justified order set enhances computational efficiency, improves model interpretability, and maintains strong approximation capability for subsequent fractional polynomial modeling tasks.

\subsubsection{Algorithm improvement based on effective order }
By replacing the original candidate set of highest orders $\mathcal{G}=\{0.1, 0.2, \dots, 0.1(N-2)\}$ with the effective order set $\mathcal{O}$, a new fractional order fitting model can be obtained.

\begin{algorithm}[H]
\caption{Fractional polynomial fitting based on effective order}\label{alg:frac_fit_bic}
\KwIn{
A collection of 3D vascular data files $\{Data_i\}$; \\
Fractional order candidate set $\mathcal{O} = \{\alpha_1, \alpha_2, \ldots, \alpha_M\}$; \\
Step size $\Delta \alpha = 0.1$
}
\KwOut{
Optimal fractional order $\alpha_{\mathrm{final}}$ for each data file; \\
Corresponding error metrics (MSE, AIC, BIC)
}

\ForEach{$Data_i \in \{Data_i\}$}{
    \tcp{(1) Data reading and parameterization}
    Read 3D data $(x, y, z)$\;
    Normalize the chord length parameter to obtain $t$\;

    \tcp{(2) Fitting and evaluation of candidate fractional-order models}
    \ForEach{$\alpha_{\max} \in \mathcal{O}$}{
        Construct order set:
        \[
            \alpha = \{\Delta \alpha, 2 \Delta \alpha, \ldots, \alpha_{\max}\};
        \]\
        Construct basis functions:
        \[
            \Phi(t) = \{t^{\alpha}, \log(t)\}, \quad 
            \text{form design matrix } A(\alpha_{\max});
        \]\
        Compute least squares coefficients:
        \[
            c_x = \arg\min \|x - A c\|_2^2, \quad
            c_y = \arg\min \|y - A c\|_2^2, \quad
            c_z = \arg\min \|z - A c\|_2^2;
        \]\
        Evaluate error indices:
        \[
            \mathrm{MSE}(\alpha_{\max}),\ \mathrm{AIC}(\alpha_{\max}),\ \mathrm{BIC}(\alpha_{\max});
        \]\
    }

    \tcp{(3) Selecting the optimal order based on BIC}
    $\alpha_{\mathrm{final}} \gets \arg\min_{\alpha_{\max} \in \mathcal{O}} \mathrm{BIC}(\alpha_{\max})$\;

    Save $\alpha_{\mathrm{final}}$ and the corresponding error metrics\;
}
\end{algorithm}

The improvement strategy based on the effective order set $\mathcal{O}$ transforms the global search that originally required calculating BIC and comparing one by one on the candidate order set $\mathcal{G}$ into a rapid discrimination on the limited set $\mathcal{O}$ only. The reduction in candidate size directly brings about an approximately linear time benefit. Assuming the original search candidate count is $|\mathcal{G}|$ and the improved candidate count is $|\mathcal{O}|$, the computational complexity of single-sample order selection decreases from $\mathcal{G}(|\mathcal{G}|\cdot C_{\mathrm{fit}})$ to $\mathcal{O}(|\mathcal{O}|\cdot C_{\mathrm{fit}}$), and the theoretical speedup ratio can be expressed as $|\mathcal{G}|/|\mathcal{O}|$. Under this setting, the size of $\mathcal{O}=\{1.1,1.5,2.0,2.7,3.4\}$ is much smaller than the original search space, thus significantly reducing the number of repeated fitting and BIC calculations in cases. At the same time, since $\mathcal{O}$ consists of main peak, secondary peak coverage terms, and stable long-tail terms, it can still maintain effective coverage for typical and complex cases. In summary, the improved algorithm achieves higher order selection efficiency without sacrificing the overall quality of order selection and fitting accuracy, thereby making it highly suitable for rapid processing of large-scale case data.

\section{\bf Experiments and Results}
\subsection{Statistical analysis of the distribution of effective orders }
Applying the approach in 2.3.1 to 379 cases, we aimed to verify the accuracy of effective order selection. The frequency distribution of $\alpha_{\max}^{*}$ exhibited a distinct "multi-peak and long tail" structure, as illustrated in Figure 5:

\begin{figure}[H]  
  \centering
  \includegraphics[width=1\textwidth]{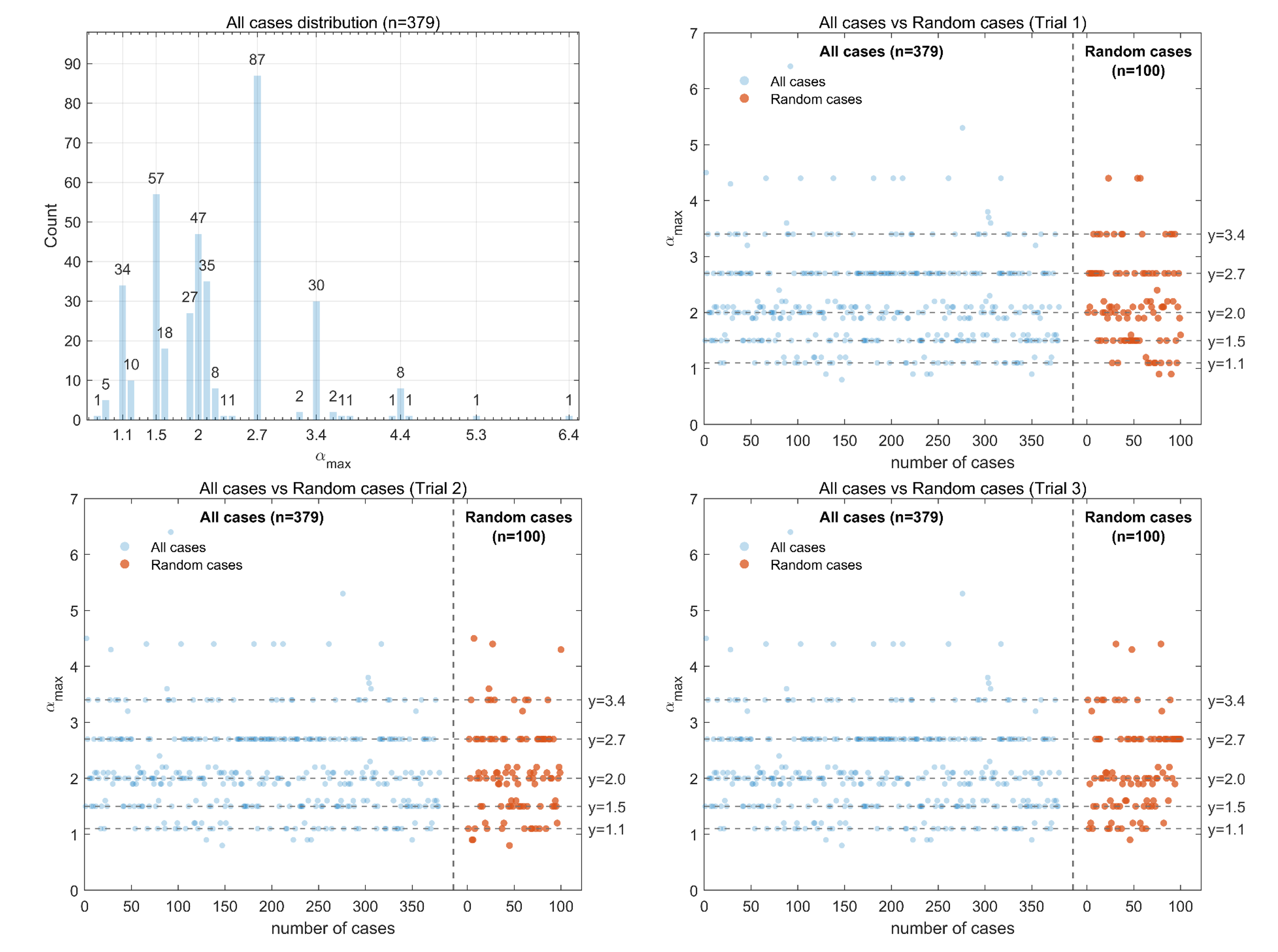} 
  \caption{Comparison of order distribution between all cases and sample cases}
  \label{fig:example}
\end{figure}

Figure 5 compares the distributions of the maximum order, $\alpha_{\rm max}$, for all 379 study subjects and three independent random samples of 100 subjects each, aiming to verify the representativeness of the samples and the generalizability of the conclusions. The full dataset exhibits a pronounced multi-peak distribution, with the vast majority of $\alpha_{\rm max}$ values concentrated in five core intervals: 1.1, 1.5, 2.0, 2.7, and 3.4. Among these, $\alpha_{\rm max} = 2.7$ occurs most frequently (23.0\%), representing the dominant maximum order in the study population. Extreme high-order cases ($\alpha_{\rm max} > 4.4$) are rare and statistically non-representative.

The results of the three independent random samples consistently fall within these core intervals, with relative frequencies closely matching those of the full dataset and distribution trends fully aligned, indicating good reliability of the sampling. Based on this analysis, subsequent investigations can focus exclusively on the five high-frequency $\alpha_{\rm max}$ values (1.1, 1.5, 2.0, 2.7, and 3.4), ensuring statistical representativeness and allowing conclusions drawn from the sample to be reliably generalized to the entire study population.


\subsection{Performance verification of improved algorithm }
\subsubsection{Operational efficiency analysis }
To verify the effectiveness of this improved strategy in practical applications, this study further designed a comparative experiment to evaluate the computational efficiency of the two methods. The experimental results showed that when processing the same scale of vascular centerline data, the traditional fractional order fitting method required 153.145 s to complete the processing of 379 cases, including multiple steps such as model construction, error calculation, and optimal order selection. However, the improved method based on effective orders completed the above steps in only 23.054 s, with the computation time reduced to $15.1\%$ of the original. This significant efficiency improvement proves that the optimization strategy of effective order selection can effectively reduce computational complexity, providing feasible technical support for real-time analysis and rapid clinical diagnosis of vascular centerlines.

\subsubsection{Analysis of fractional polynomial fitting results }
Furthermore, this article conducted comparative validation experiments on 379 samples to evaluate the differences in MSE between the original method and the improved method. The results are shown in Figure 6.

\begin{figure}[H]  
  \centering
  \includegraphics[width=1\textwidth]{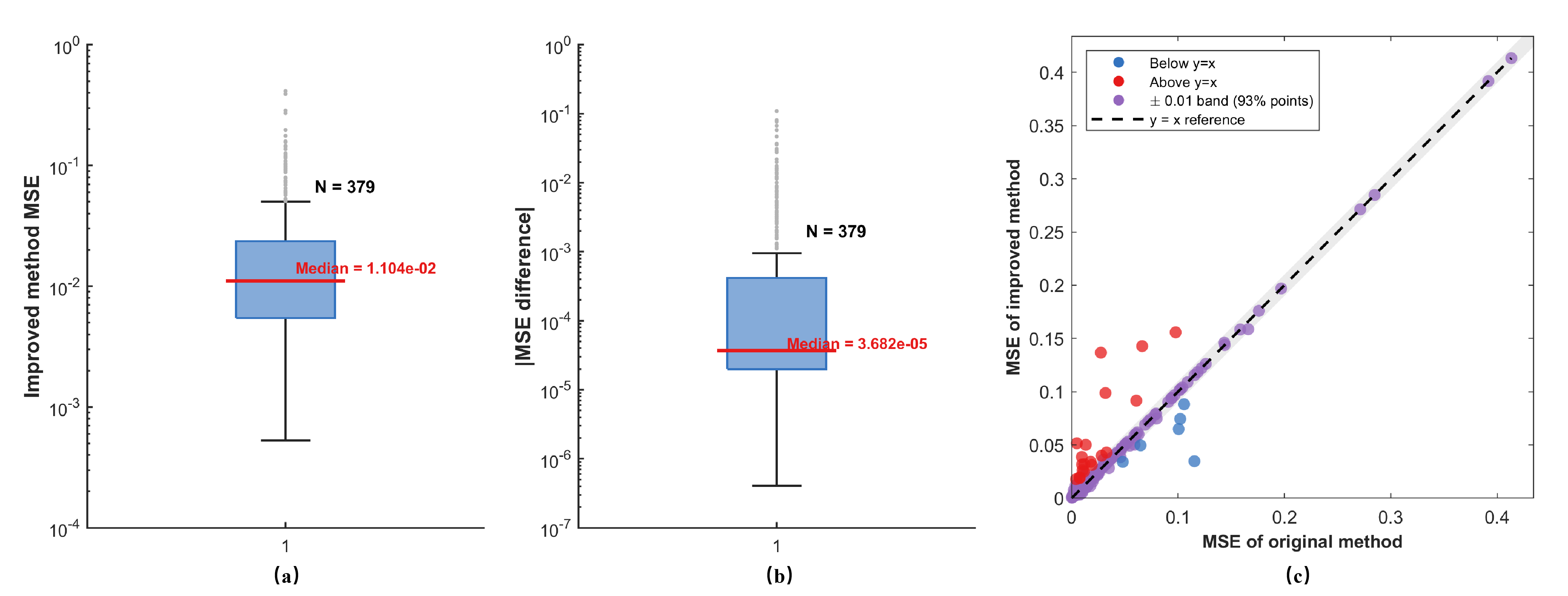} 
  \caption{Comparison of MSE performance between the improved method and the original method. (a) Improved MSE distribution; (b) Absolute MSE differences; (c) MSE scatter plot. Purple points fall within the ±0.01 error band.}
  \label{fig:example}
\end{figure}

As shown in Figure 6, the improved method exhibits significant advantages in terms of MSE distribution and performance comparison. As shown in Figure 6(a), in the set containing 379 test samples, the median MSE of the improved method is $1.104 \times 10 ^ {-2} $, and the MSE of most samples is concentrated in the range of $10 ^ {-3} $ to $10 ^ {-1} $, with only a few high MSE outliers, demonstrating the stability and reliability of the overall performance. Figure 6(c) further compares the MSE performance of the improved method and the original method through a scatter plot. Using y=x as the reference line, about $ 93\%$ of the sample points (purple) are located within the error band of 0.01 pm, indicating that the error level of the improved method is highly similar to that of the original method on most samples. The blue dots indicate that the MSE of the improved method is lower than that of the original method, while the red dots indicate that the MSE is higher than that of the original method. The overall point distribution is closely centered around the reference line. Based on the analysis of the absolute difference box plot (Figure 6(b)), it can be seen that the MSE differences of the vast majority of samples are concentrated in the lower range, further indicating that the improved method can maintain similar fitting accuracy as the original method overall, while significantly reducing the computational cost of order search in fractional polynomial fitting, demonstrating good robustness and reliability.

To visually demonstrate the effectiveness of fractional-order polynomial fitting based on the effective order, this article selects several typical cases for analysis. In these cases, the highest order of the fractional-order polynomial should be selected from the set of effective orders, while the highest order of the integer-order polynomial is rounded up to the nearest fractional order. Figure 7 shows the fitting results for different typical sample types.

\begin{figure}[H]  
  \centering
  \includegraphics[width=1\textwidth]{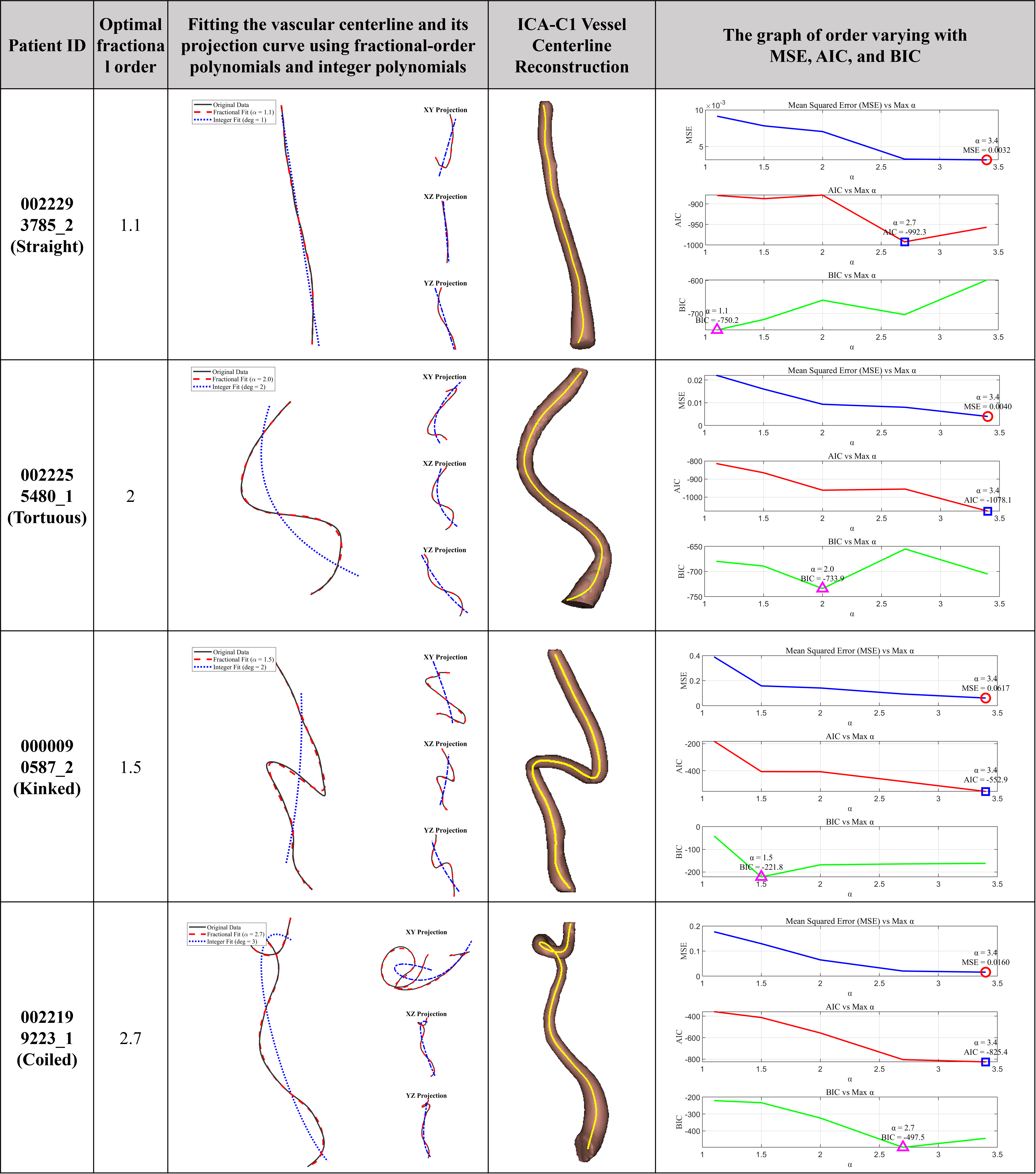} 
  \caption{Comparison of fitting effects between fractional order and integer order}
  \label{fig:example}
\end{figure}

In summary, the improved strategy based on the effective order set proposed in this study was rigorously validated in an experiment involving 379 samples. This method not only reduces the computational load of order search, improves operational efficiency, but also achieves very low fitting error while maintaining high accuracy and consistency.

\subsubsection{Analysis of robustness}
To evaluate the noise robustness performance of the proposed method, Gaussian noise with different intensities(noise intensities of 0, 0.01, 0.05, and 0.1) was introduced on the overall sample of 379 cases, and the distribution of prediction errors under various noise conditions was compared, as shown in the figure 8.

\begin{figure}[H]  
  \centering
  \includegraphics[width=1\textwidth]{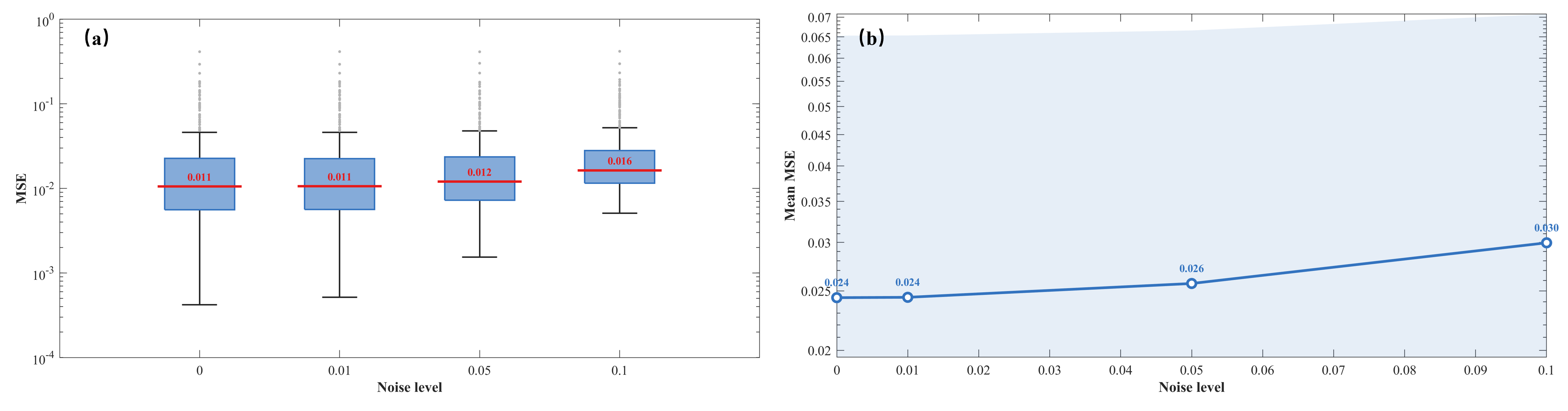} 
  \caption{Robustness analysis of the proposed model under different noise levels: (a)MSE distribution under different noise levels; (b)Trend of Mean MSE under Different Noise Levels}
  \label{fig:example}
\end{figure}

The experimental results indicate that as the noise intensity gradually increases, the mean squared error (MSE) of the model shows an upward trend to some extent, but the changes in the median and dispersion of the error distribution are relatively limited. At higher noise levels, the error distribution remains relatively concentrated, with no significant performance degradation observed.

The above results demonstrate that the proposed method exhibits strong noise robustness and consistent stability across the sample set, indicating its reliability and effectiveness for vascular geometric analysis.

\subsection{Clinical tortuosity parameter analysis}
Blood vessel centerline data extracted from medical imaging are typically discrete points, which can be converted into a continuous representation via fractional polynomial fitting. From this, curvature $\kappa$ and torsion $\tau$ can be numerically computed (Formula 2.1) to quantify geometric bending and spatial twisting of the vessels. Several representative tortuosity indicators are defined based on these calculations.

Total Curvature (TC) is defined as
$
TC = \int \kappa(s) \, ds = \int_a^b \kappa(t) \cdot \| \mathbf{r}'(t) \| \, dt,
$
which measures the overall degree of bending along the vessel.
Similarly, Total Torsion (TT) describes the overall spatial twisting of the vessel and is given by
$
TT = \int \tau(s) \, ds = \int_a^b \tau(t) \cdot \| \mathbf{r}'(t) \| \, dt.
$
To eliminate the influence of vessel length, the length-normalized total curvature (TC/L) and length-normalized total torsion (TT/L) are introduced:
$
\frac{TC}{L} = \frac{\int_{t_0}^{t_N} \kappa(t) \|\mathbf{r}'(t)\| \, dt}{L}, \quad
\frac{TT}{L} = \frac{\int_{t_0}^{t_N} |\tau(t)| \|\mathbf{r}'(t)\| \, dt}{L},
$
which allow standardized comparisons between vessels of different lengths. Finally, the Tortuosity Index (TI), reflecting the overall degree of vessel tortuosity, is defined as
$
TI = \frac{L}{D}.
$
Together, these indicators provide a comprehensive quantification of vascular bending and twisting characteristics.

Among these, TC and TT, as global integration indicators, characterize the bending and twisting characteristics of blood vessels at the overall scale. TC/L and TT/L facilitate standardized comparisons between vessels of different lengths\textsuperscript{\cite{Poirier2025Standardizing, Bullitt2003Measuring, zhang2021application}}. The calculation results of each indicator are shown in Figure 9.

\begin{figure}[H]  
  \centering
  \includegraphics[width=1\textwidth]{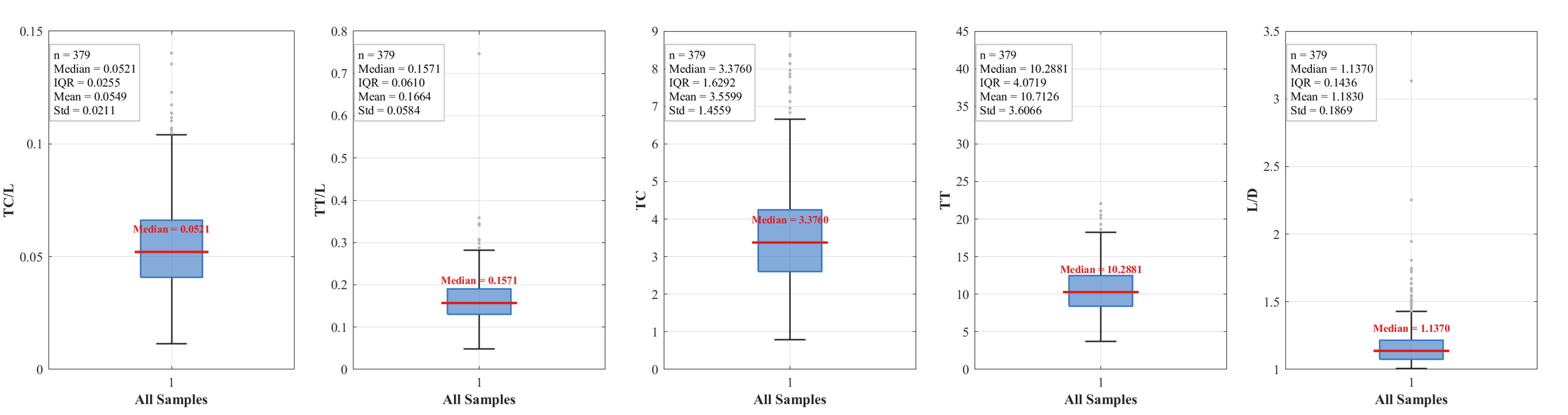} 
  \caption{Distribution of parameters for vascular tortuosity}
  \label{fig:example}
\end{figure}

The figure above illustrates the distribution of 379 samples across five tortuosity parameters (TC, TT, TC/L, TT/L, and L/D). Overall, the distributions are relatively concentrated, though individual variability exists. As global integral indicators, TC and TT exhibit greater dispersion and a mild right-skewed distribution, suggesting that a subset of vessels display pronounced bending or spatial twisting. After arc length normalization, the distributions of TC/L and TT/L converge markedly, with reduced interquartile ranges, indicating that normalization effectively mitigates the influence of vessel length differences and enhances comparability across samples. The tortuosity index L/D is primarily concentrated between 1.0 and 1.3, suggesting that most vessels exhibit mild to moderate tortuosity, with only a few samples showing high tortuosity. Overall, the distributions of these indicators appear reasonable and provide a reliable reflection of the statistical characteristics of vascular geometric features.

\subsection{Effective-Order Fractional Polynomial Vascular Prediction}
During CTA data acquisition, vascular images often suffer from non-visualization caused by vessel occlusion, severe stenosis, insufficient contrast enhancement, motion artifacts, or other imaging-related factors. To overcome these challenges, particularly the blurred visualization or missing data in the middle sections of blood vessels, the fractional-order polynomial fitting model introduced in the previous section is employed as the core reconstruction framework. The effective order of the model is first determined using the BIC, ensuring an appropriate balance between fitting accuracy and model complexity. For the specific task of vascular occlusion or missing-segment prediction, boundary constraints are further incorporated into the least-squares estimation to enforce continuity at both sides of the incomplete region. This strategy enables more accurate reconstruction of vascular spatial trajectories while preserving the geometric shape and directional consistency of the vessel. Consequently, it improves the reliability of subsequent quantitative analysis and provides a robust solution for incomplete or partially visualized vascular datasets.

In this study, a simulation strategy was used to model missing vessels in the middle segment. First, the three-dimensional centerline coordinates $(x,y,z)$ of each vessel were extracted, and the trajectory was parameterized and normalized to reduce the influence of different vessel lengths. Then, approximately 10\% of the middle portion was removed artificially, while the proximal and distal segments were retained as known data for model fitting, and the missing middle segment was used as the prediction target.The detailed implementation procedure is summarized in the following algorithm.

\begin{algorithm}[H]
\caption{Fractional-Order Trajectory Reconstruction via Constrained Least Squares}
\label{alg:frac_recon}
\KwIn{
Trajectory set $\mathcal{D}=\{Data_i\}$; candidate orders $\mathcal{A}=\{1.1,1.5,2,2.7,3.4\}$; boundary weight $w$
}
\KwOut{
Best model, reconstructed segment, MSE, NMSE
}

\ForEach{$Data_i\in\mathcal{D}$}{

    Load trajectory $(x,y,z)$ with length $N$\;

    Compute normalized arc-length parameter $t$\;

    Remove the middle $10\%$ samples and divide data into known part $(t_k)$ and missing part $(t_m)$\;


    \ForEach{$\alpha_{\max}\in\mathcal{A}$}{

        Generate fractional orders:
        $\alpha=0.1:0.1:\alpha_{\max}$\;

        Build basis matrix
        $
        A=[1,\ t_k^{\alpha_1},\dots,t_k^{\alpha_m},\log(t_k)]
        $

         Build missing-segment basis matrix
         $
         A_m=[1,\ t_m^{\alpha_1},\dots,t_m^{\alpha_m},\log(t_m)]
         $
         
        Define boundary basis vectors at the two gap endpoints:
        \[
        A_L=\phi(t_L),\quad A_R=\phi(t_R)
        \]

        \tcp{Constrained least squares}
        Enforce endpoint continuity by adding weighted boundary equations:
        \[
        A_{\mathrm{aug}}=
        \begin{bmatrix}
        A\\
        wA_L\\
        wA_R
        \end{bmatrix},\quad
        b_x=
        \begin{bmatrix}
        x_k\\
        wx(t_L)\\
        wx(t_R)
        \end{bmatrix}
        \]
        similarly for $b_y,b_z$\;

        Solve the weighted least-squares problem:
        \[
        c_x=\arg\min\|A_{\mathrm{aug}}c-b_x\|^2,\quad
        c_y=\arg\min\|A_{\mathrm{aug}}c-b_y\|^2,\quad
        c_z=\arg\min\|A_{\mathrm{aug}}c-b_z\|^2
        \]

        Compute BIC, and retain the model with the minimum BIC and its coefficients $c_x,c_y,c_z$\;
    }

    Reconstruct missing segment:
    \[
    \hat{x}_m=A_mc_x,\quad
    \hat{y}_m=A_mc_y,\quad
    \hat{z}_m=A_mc_z
    \]

    Compute MSE and NMSE\;

    Save results\;
}
\end{algorithm}

In the modeling phase, a fractional-order fitting model based on the effective order is employed. Using known data from the anterior and posterior ends of the blood vessel, the model captures the overall spatial trajectory of the vessel. Once the optimal model expression is obtained, it is extrapolated to the missing interval in the middle section, enabling prediction and reconstruction of the spatial trend in non-visualized regions of the vessel. The predicted trajectory is then compared with the actual vascular path. Reconstruction performance is quantitatively assessed using the mean square error (MSE) and normalized mean square error (NMSE).


Based on the above process, a stable and effective prediction of the morphology and spatial orientation of blood vessels at the location of missing visualization in the middle segment was achieved by employing fractional-order polynomials based on the effective order. This approach not only accurately reconstructs the overall vascular trajectory, but also maintains high fidelity to the actual anatomical structure, demonstrating strong robustness and generalizability for incomplete or partially visualized vascular data.

To visually demonstrate the predictive performance of the fractional order model proposed in this study under the condition of missing visualization in the middle segment of blood vessels, this paper selects several representative blood vessel samples and visually compares and analyzes their spatial trajectory prediction results. As shown in Figure 11, the complete trajectory of the original blood vessels, the state of missing imaging, and the predicted reconstruction results based on the model proposed in this paper are presented separately.

\begin{figure}[H]  
  \centering
  \includegraphics[width=1\textwidth]{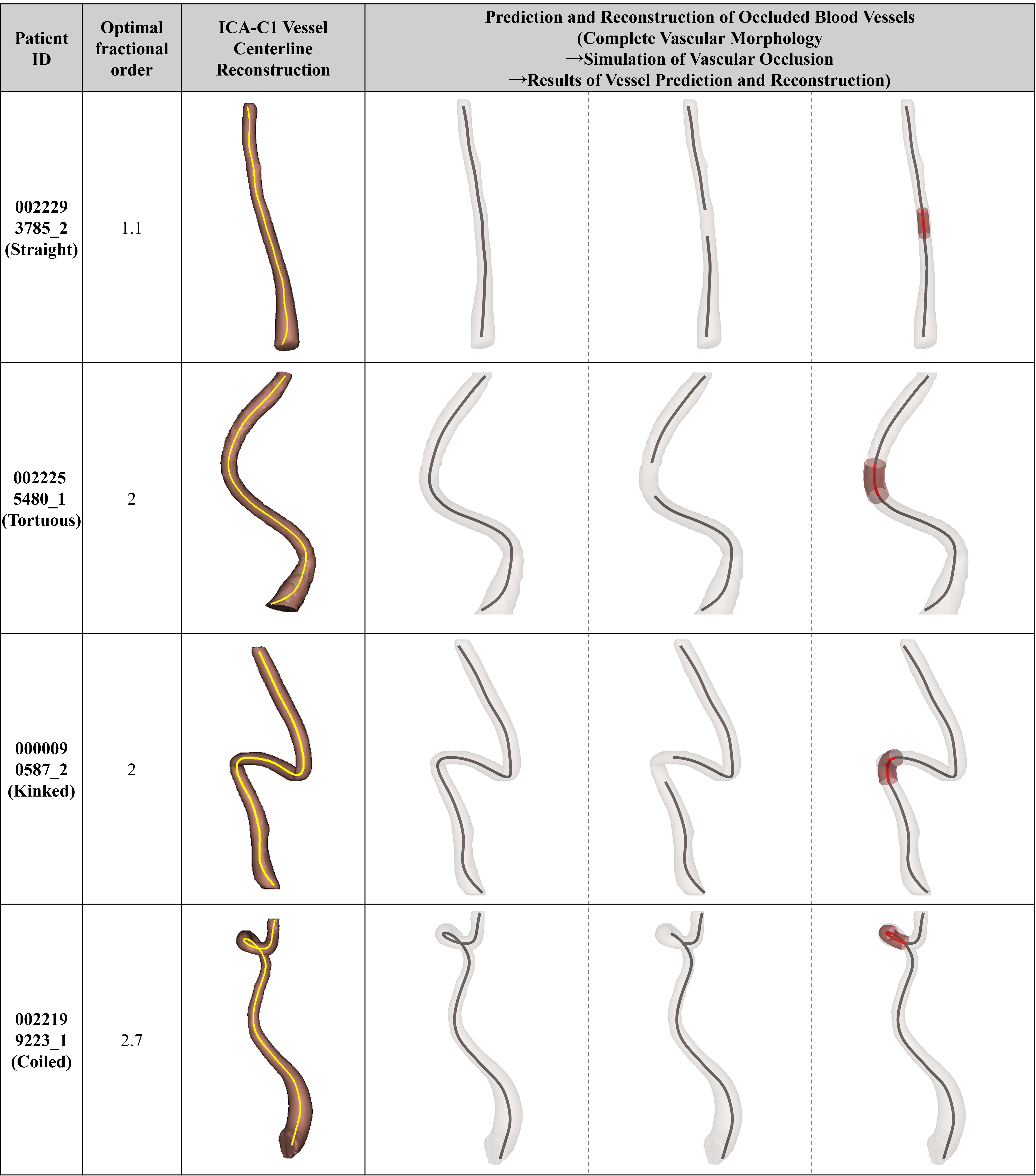} 
  \caption{Effect diagram of fractional order polynomial vascular prediction based on effective order}
  \label{fig:example}
\end{figure}

Figure~10(a) shows the complete true trajectory of the blood vessel used for verification. Figure~9(b) illustrates the simulated missing-visualization condition, where the middle segment of the vessel is removed, leaving only the observable portions at both ends. Figure~9(c) presents the reconstruction results predicted by the fractional-order model. In the figure, the gray curve represents the complete original trajectory, the green curve denotes the true trajectory within the predicted interval, and the red semi-transparent tubular structure represents the model prediction. To further validate the applicability and stability of the proposed method under different vascular morphological conditions, the reconstruction results are categorized and presented according to vascular geometric characteristics. Specifically, representative cases are selected from three typical categories: approximately straight vessels, mildly curved vessels, and highly tortuous vessels, and are visualized and analyzed separately. For each category, the original complete trajectory, the observed trajectory with the middle segment removed, and the reconstructed trajectory obtained by the proposed model are presented for comprehensive comparison. 

\begin{figure}[H]  
  \centering
  \includegraphics[width=1\textwidth]{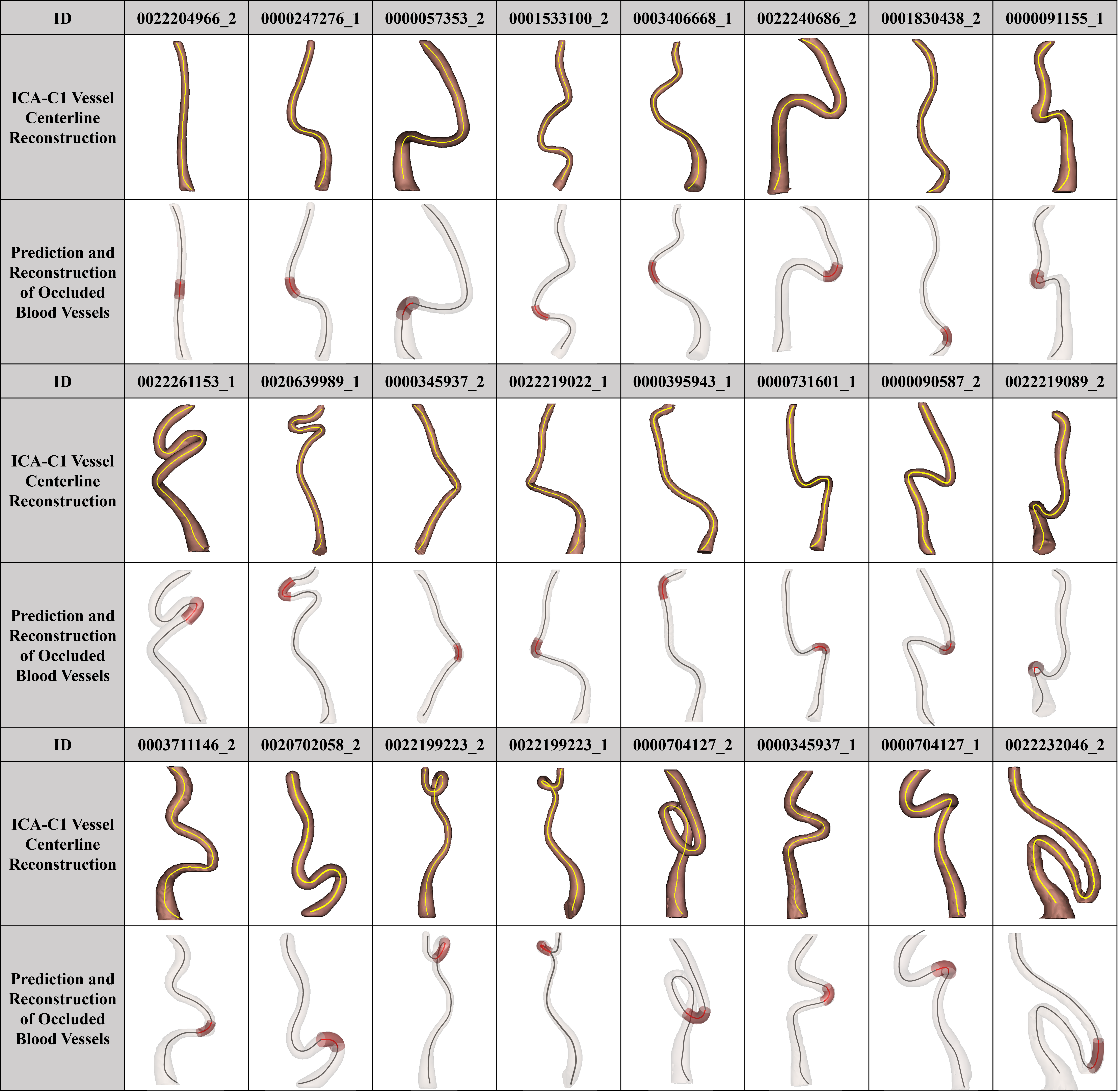} 
  \caption{Reconstruction results for vascular trajectories.}
  \label{fig:example}
\end{figure}

For both straight and tortuous vascular samples, the proposed method demonstrates exceptionally high reconstruction accuracy. The predicted trajectories are in close agreement with the ground truth across all cases, preserving both directional consistency and geometric smoothness. Owing to the relatively simple vessel structures and limited curvature variation, the fractional-order model is able to accurately capture the global trend of the vessels using only boundary information. These results confirm that the model performs reliably under low-complexity conditions and is capable of achieving near-perfect reconstruction for straight and gently curved vessel segments.


For coiled vascular samples, the reconstruction task becomes more challenging due to the increased geometric complexity and rapid directional changes. Nevertheless, the predicted trajectories still capture the overall structural trend of the vessels and maintain global continuity. Although some local discrepancies may occur in regions with sharp curvature variations, the model does not exhibit severe instability or divergence. These results demonstrate that the proposed method retains a certain level of robustness even under highly complex morphological conditions.


For kinked vascular samples, where abrupt changes in direction are present, the reconstruction performance shows a moderate decline compared to smoother vessel types. The model is still able to approximate the overall trajectory; however, localized errors tend to increase near sharp turning points. This is mainly due to the difficulty in accurately capturing sudden geometric transitions using a globally fitted model. Despite these challenges, the predicted trajectories remain free from significant oscillations, and the overall vessel topology is well preserved. This suggests that the proposed method maintains stability even in the presence of discontinuous curvature variations. Subsequently, statistical analysis is performed on the prediction results of all samples, as summarized in the table below.
Subsequently, statistical analysis is performed on the prediction results of all samples, as summarized in the table below.


\renewcommand{\arraystretch}{1.2}
\setlength{\arrayrulewidth}{1.0pt}
\begin{center}
\captionsetup{aboveskip=2pt, belowskip=2pt}
\captionof{table}{Summary of Vascular Tortuosity Parameters}
\label{tab:vascular_tortuosity_parameters}

\begin{tabular}{|c|c|c|c|c|c|c|}
\hline
\rowcolor{gray!40}
\textbf{Index} & \textbf{Median} & \textbf{Mean} & \textbf{75th percentile} & \textbf{90th percentile} & \textbf{80th percentile cutoff} & \textbf{90th percentile cutoff} \\
\hline
\textbf{NMSE} & 0.31\% & 1.01\% & 0.73\% & 1.68\% & 0.82\% & 1.68\% \\
\hline
\textbf{MSE}  & 0.01 & 0.02 & 0.02 & 0.04 & 0.02 & 0.04 \\
\hline
\end{tabular}
\end{center}

Visualize the above data as shown in the following figure.

\begin{figure}[H]  
  \centering
  \includegraphics[width=1\textwidth]{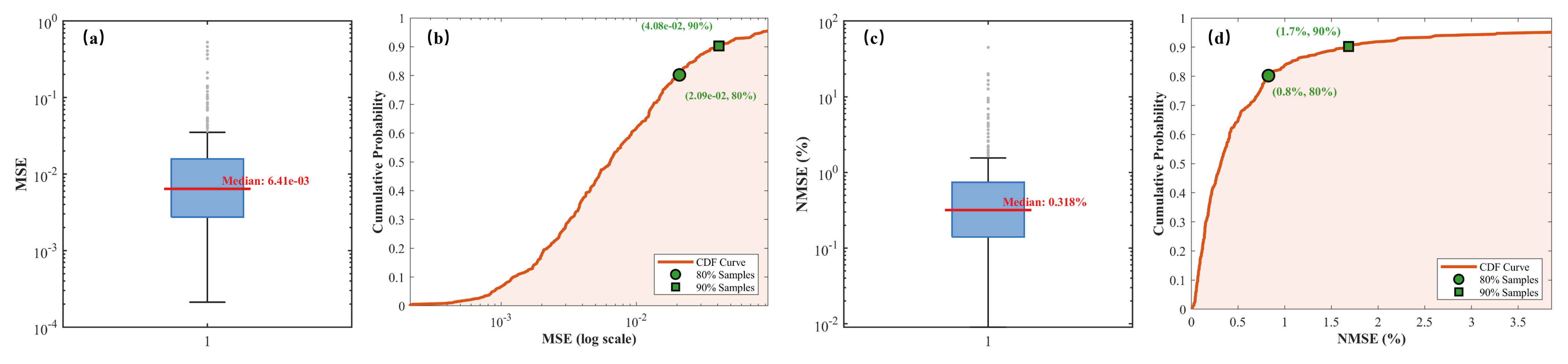} 
  \caption{Analysis of vascular prediction results based on error metrics: (a) Distribution of NMSE; (b) Cumulative distribution of NMSE.\\(c) Distribution of MSE; (d) Cumulative distribution of MSE.}
  \label{fig:example}
\end{figure}

Firstly, Figure 12(a) presents the boxplot distribution of NMSE for all vascular samples. It can be seen that the NMSE of most samples is concentrated in the lower range, with a median of only $0.31\%$ and a 75th percentile of $0.73\%$, indicating that the model has high prediction accuracy on most vascular segments. Meanwhile, there are still a few outliers with high NMSE in the boxplot, and their maximum values are significantly higher than the overall distribution. These types of samples usually correspond to areas with highly curved blood vessels, complex branches, or drastic geometric changes. In the absence of imaging information in the middle segment, the prediction process faces stronger nonlinearity and uncertainty, leading to an increase in errors.

To further quantitatively characterize the overall performance, Figure 12(b) shows the cumulative distribution function (CDF)\textsuperscript{\cite{Hammerla2013On}} of NMSE. The results showed that about $80\%$ of the samples had NMSE below $0.82\%$, and about $90\%$ had NMSE below $1.68\%$. This result indicates that despite the presence of a small number of high error samples, the fractional order model can still stably reconstruct the vascular trend in the middle segment without visualization in the vast majority of cases, demonstrating strong robustness and ability to generalize vascular trends.

In addition, Figures 12(c) and (d) show the distribution and cumulative distribution of MSE. MSE is $0.02$, and the 75th percentile and 90th percentile are $2.0 \times 10^{-2}$ and $4.0 \times 10^{-2}$, respectively. CDF results show that about $80\%$ of the samples have MSE below 0.02 and about $90\%$ of the samples have MSE below 0.04, further verifying the stable predictive ability of the model in the sense of absolute error.

The comprehensive box plot, CDF curve, and statistical indicators can demonstrate that the proposed fractional order prediction model can effectively capture the continuity and subtle structural features of blood vessel orientation under the condition of missing visualization in the middle segment. Although there are still some errors in highly nonlinear and complex geometric structures, the overall model has achieved accurate and reliable predictions in most vascular segments, providing effective methodological support for the reconstruction of missing vascular segments.

\section{Conclusion}

This article proposes a fractional polynomial fitting method based on feature order for the geometric modeling of the centerline of the ICA – C1 segment blood vessels, and systematically validates it on large-scale clinical data. The research results indicate that the constructed fractional order polynomial model can flexibly express complex vascular morphology, effectively reduce oscillation and overfitting phenomena, and has higher fitting accuracy compared to traditional integer order polynomials. By introducing a BIC based order automatic selection mechanism and the concept of "effective order", the algorithm achieves order adaptive optimization and demonstrates significant advantages in computational efficiency and stability.

The experimental results show that the improved algorithm has better fitting accuracy, anti noise ability, and computational efficiency than traditional methods in 379 clinical cases. The improved algorithm obtains extremely small fitting errors and can more accurately characterize the complex spatial morphology features of the ICA – C1 segment. Under the conditions of adding noise of different intensities of 0, 0.01, 0.05, and 0.1, stable and good fitting can still be achieved. At the same time, the algorithm greatly optimized the computational complexity, and the running time of the optimized algorithm decreased from 153.145 s to 23.054 s. In addition, in the experiment of predicting missing blood vessel segments, the model performed stably, and the prediction error of most cases was low. The normalized mean square error (NMSE) of 90\% of cases was controlled below 1.68\%, indicating that the method has good generalization and reliability. Statistical analysis based on curvature and torsion indicates that this method can provide quantitative references for vascular classification, risk assessment, and preoperative planning, and has potential clinical application value.

In the future, this method can be further extended to more complex vascular networks and multi center datasets to improve its universality and clinical applicability. At the same time, it can be combined with hemodynamic simulation to explore the relationship between vascular geometric features and blood flow characteristics, providing more comprehensive indicator support for risk prediction of cerebrovascular diseases. In addition, integrating this method with surgical navigation systems or intelligent diagnostic platforms is expected to achieve automated centerline modeling and morphological analysis, providing technical support for clinical decision-making and preoperative planning.

In summary, the fractional order polynomial fitting method proposed in this article demonstrates high stability, accuracy, and efficiency in modeling and morphological analysis of vascular centerlines, providing a feasible and scalable technical solution for quantitative analysis and intelligent decision-making of cerebrovascular diseases.

\vskip3mm

\noindent
{\bf Conflicts of interest:} The authors declare no conflict of interest.

\end{document}